\documentclass{aa}  

\usepackage{graphicx}
\usepackage{longtable}
\usepackage{tabularx, blindtext}
\usepackage{lscape}
\usepackage[varg]{txfonts}
\usepackage{lmodern}
\usepackage[breaklinks]{hyperref}
\newcommand{\teff}{\ensuremath{T_\mathrm{eff}}}
\newcommand{\logg}{\ensuremath{\log g}}
\newcommand{\kms}{$\rm km\,s ^{-1}$}

\usepackage{color}

\begin{document}

   \title{Sulfur abundances in three Galactic clusters: \\Ruprecht 106, Trumpler 5, and Trumpler 20 \thanks{This paper is based on data collected with the Very Large Telescope (VLT) at the European Southern Observatory (ESO) on Paranal, Chile (ESO Program ID 69.D-0642, 098.D-0227, 099.D-0505, 0100.D-0262, 188.B-3002).}}

   \author{F. Lucertini\inst{1,2}, L. Monaco\inst{1}, E. Caffau\inst{3}, A. Mucciarelli\inst{4,5}, S. Villanova\inst{6}, P. Bonifacio\inst{3}, L. Sbordone\inst{2}} 

   \institute{Departamento de Ciencias Fisicas, Faculdad de Ciencias Exactas, Universidad Andres Bello, Av. Fernandez Concha 700, Las Condes, Santiago, Chile
   \and
   ESO - European Southern Observatory, Alonso de Cordova 3107, Vitacura, Santiago, Chile
   \and
   GEPI, Observatoire de Paris, Universit\`e PSL, CNRS, Place Jules Janssen, 92195 Meudon, France
   \and
   Dipartimento di Fisica e Astronomia, Università degli Studi di Bologna, Via Gobetti 93/2, I-40129 Bologna, Italy
        \and    
        INAF - Osservatorio di Astrofisica e Scienza dello Spazio di Bologna, Via Gobetti 93/3, I-40129 Bologna, Italy
        \and
        Departamento de Astronomia, Casilla 160-C, Universidad de Concepcion, Cooncepci\'{o}n, Chile
  }

   \date{Received 1 August 2022; Accepted 5 October 2022}

 
  \abstract
   {Sulfur (S) is one of the lesser-studied $\alpha$-elements. Published investigations of its behavior have so far focused on local stars, and only a few clusters of the Milky Way have been considered to study this topic.
   We aim to study the S content of the globular cluster Ruprecht\,106 ---which has never before been studied for  this purpose, but is known to present low levels of the [$\alpha$/Fe] abundance
ratio--- and the open cluster Trumpler\,5. The only star studied so far in Trumpler\,5 shows an unexpectedly low abundance of S.
}
   {With this work, we aim to provide the first S abundance in Ruprecht\,106 and to investigate the S content of Trumpler\,5 with a larger sample of stars. The open cluster Trumpler\,20 is considered as a reference object.}
   {We performed a standard abundance analysis based on 1D model atmospheres in local thermodynamical equilibrium (LTE) and on high-resolution and high-signal-to-noise-ratio UVES-slit and UVES/FLAMES spectra. We also applied corrections for nonLTE. The metallicities of the targets were obtained by studying equivalent widths. Sulfur abundances were derived from multiplets 1, 6, and 8 by spectrosynthesis.}
   {We find that the metallicities of Ruprecht\,106 and Trumpler\,5 are [Fe/H]= --1.37$\pm$0.11 and [Fe/H]= --0.49$\pm$0.14, respectively. Ruprecht\,106 is less S-rich than the other Galactic clusters at similar metallicity. The low S content of Ruprecht\,106, [S/Fe]$_{\rm NLTE}$= --0.52$\pm$0.13, is consistent with its shortage of $\alpha$-elements. This supports an extra-galactic origin of this cluster. We obtained a new and more robust S content value of Trumpler\,5 of about [S/Fe]$_{\rm NLTE}$= 0.05$\pm$0.20. According to our results, Trumpler\,5 follows the trend of the Galactic disk in the [S/Fe]$_{\rm LTE}$ versus [Fe/H] diagram. Our results for Trumpler\,20, of namely [Fe/H]= 0.06$\pm$0.15 and [S/Fe]$_{\rm NLTE}$= --0.28$\pm$0.21, are in agreement with those in the literature.} 
   {}

   \authorrunning{F. Lucertini et al.}
   \maketitle

\section{Introduction}
The $\alpha$-elements (O, Ne, Mg, Si, S, Ar, Ca and Ti) are recognized by the scientific community as a powerful tool for reconstructing the chemical evolution of stellar populations and galaxies. They are produced by successive fusions of helium nuclei ($\alpha$ particles) in massive stars ($M_* > 8 M_{\odot}$), which explode as type-II supernovae (SNe II) at the end of their life and eject their material into the interstellar medium (ISM). On the other hand, type-Ia SNe (SNe Ia) mainly release iron-peak elements (Cr, Mn, Fe, Co, Ni Cu, Zn) in the ISM over longer timescales. The time delay between the explosion of SNe II and SNe Ia, and consequently the [$\alpha$/Fe] ratio, provide crucial clues with which to infer the star formation history and the evolution of a system. 
For these kinds of investigations, the works in the literature usually skip the analysis of sulfur (S) in favor of other $\alpha$-elements. Consequently, our knowledge about S is far from complete.

In agreement with the $\alpha$-elements behavior, the [S/Fe] versus [Fe/H] diagram for Milky Way (MW) stars shows a plateau around [S/Fe]$\sim$0.4 at low [Fe/H], followed by a decrease in [S/Fe] with increasing metallicity until reaching [S/Fe]=0 at about [Fe/H]=0 \citep{clegg1981, francois1987, nissen2007}.
This trend is confirmed by studies of halo \citep{francois1988, nissen2004, caffau2010} and disk \citep{chen2002, ryde2006, takeda2011, duffau2017} Galactic stars, but the situation is less clear in the low-metallicity regime.
The [S/Fe] value obtained by \cite{israelian2001} constantly increases as the metallicity decreases, until $\sim0.7-0.8$ dex at $-2.3<$[Fe/H]$<-1.9$. These authors  explain this value by proposing that hypernovae contributed to the nucleosynthesis of elements in early galaxies. The time-delay deposition of iron into the ISM could be another explanation \citep{ramaty2001}. Instead, \cite{caffau2005a} found a bimodal behavior of [S/Fe] (both stars with [S/Fe]$\sim$0.4 and higher values) at metallicities lower than [Fe/H]$<-1.0$.
On the other hand, the S trend presented by \cite{matrozis2013} in the metallicity range $-2.5<$[Fe/H]$<-1.0$ is characterized by a plateau at [S/Fe]$\sim 0.4$, which is typical of $\alpha-$elements.
Finally, \cite{spite2011} investigated S in extremely metal-poor stars ($-3.2<$[Fe/H]$<-2.0$) and found no correlation between [S/Fe] and [Fe/H].

Recently, \cite{griffith2020} and \cite{lucertini2022} confirmed that the behavior of S resembles that of $\alpha$-elements within the Galactic bulge. However, these authors found conflicting results comparing sulfur abundances (referred to hereafter as A(S)) of disk and bulge stars in the MW. \cite{griffith2020} found that the disk and the bulge have similar S trends, while the Galactic bulge is S-rich with respect to both the thick and the thin disk according to \cite{lucertini2022}. 

So far, A(S) have only been obtained for a few globular clusters (GCs) and open clusters (OCs). \cite{sbordone2009} estimated the first A(S) in subgiant stars of NGC\,6752 and 47\,Tuc. The GC NGC\,6397 was analyzed by \cite{koch&caffau2011}. \cite{kacharov2015} obtained A(S) for the GCs M\,4, M\,20, and M\,30. The most inconsistent result was obtained by \cite{caffau2014} for the OC Trumpler 5. Indeed, the only analyzed star of this object was found to be
characterized by a low A(S) value. In the same work, the authors studied the GC M\,4, and the OCs NGC\,2477 and NGC\,5822. Finally, \cite{duffau2017} made a significant contribution, studying five new GCs and 16 OCs. All these authors concluded that S behaves like the other $\alpha$-elements.

Considering that the MW hosts more than 150 
GCs\footnote{\url{https://people.smp.uq.edu.au/HolgerBaumgardt/globular/}} and less than one-tenth have been taken into account for the investigation of S, it is clear that more effort should be devoted in this subject. The aim of the present study is to increase the numbers of clusters and stars of known A(S) content.
In particular, we are presenting Fe and S abundances for the GC Ruprecht 106 (Rup\,106) and the OCs Trumpler 5 (Tr\,5) and Trumpler\,20 (Tr\,20).

The paper is structured as follows: in section 2, we summarize the relevant information gathered so far about the clusters considered in this study and our reasons for analyzing them.
Observational data are described in section 3. Section 4 presents our data analysis and chemical abundance estimations. In section 5, we discuss our results and compare them with the literature. Finally, our conclusions are summarized in section 6.

\section{The targets}
\subsection{Ruprecht\,106}
Rup\,106 is a GC located in the MW halo ($12^h 38^m 40.2^s , -51^{\circ}09'01''$) at 21.2 kpc from the Sun and 18.5 kpc from the Galactic center \citep{Harris2010}. At an apparent visual distance modulus (m-M)$_V$=17.25, its reddening is E(B-V)=0.20 \citep{Harris2010}. With an age of 12 Gyr \citep{dacosta1992, VandenBerg2013, frelijj2021}, Rup\,106 is younger than the bulk of the Galactic GCs \citep{buonanno1993, dotter2011}. The works in the literature agree on the metal-poor nature of Rup\,106, and provide values of between [Fe/H]= -1.66 \citep{dacosta1992, francois1997, Harris2010} and 
[Fe/H]= -1.45 \citep{pritzl2005, frelijj2021}. Rup\,106 is considered the first convincing example of a single stellar population GC \citep{villanova2013, dotter2018, frelijj2021}.
This assessment is supported by the absence of an Na-O anti-correlation \citep{buonanno1990, villanova2013, frelijj2021}.
Moreover, the color--magnitude diagram (CMD) of Rup\,106 shows a very narrow red-giant branch (RGB; \citealt{dotter2018, frelijj2021}).
From a chemical point of view, Rup\,106 is over-deficient in $\alpha$-elements (\citealt{brown1996}, \citeyear{brown1997}, \citealt{villanova2013}, \citealt{francois2014}). 
In particular, its chemical composition is consistent with Local Group (LG) dwarf galaxies, and it is representative of nucleosynthesis processes different from those that have taken place in the majority of the Galactic halo and nearby clusters \citep{villanova2013}. For these reasons, it has been proposed that Rup\,106 formed outside the Milky Way, and was accreted by our galaxy \citep{lin1992, brown1996, francois2014}.

 Rup\,106 has never before been considered for the study of S. In this work, we provide the first A(S) in this cluster. In particular, we investigated whether or not Rup\,106 is characterized by a low content of S, in agreement with the behavior of the other $\alpha$-elements. 

\subsection{Trumpler\,5}
Tr\,5 is an OC located in the MW disk (RA, DEC)=(6$^h$36$^m$42$^s$, +09$^\circ26'00''$), at 3.1 $\pm$ 0.1 kpc from the Sun \citep{kim2009}. \cite{piatti2004} estimated an angular radius of 7.7 arcmin (5.4 pc) for this object. The distance modulus and the reddening of Tr\,5 are (m-M)$_V$=12.4 and E(B-V)=0.58 \citep{kaluzny1998, kim2009, donati2015}, respectively. Due to its old age (4 Gyr, \citealt{kaluzny1998, donati2015}), Tr\,5 is an ideal laboratory for studying the formation and early evolution of the Milky Way disk.
Indeed, old OCs constitute only $\sim 15\%$ of those known so far \citep{dias2002, cantat2020}, and are probes of the structure and chemical distribution of the Galactic disk.  
Tr\,5 is a metal-poor OC with [Fe/H]= $-0.4 \pm $0.1 \citep{kaluzny1998, piatti2004, kim2009, donati2015}, and is characterized by solar abundance ratios \citep{donati2015, monaco2014}. Moreover, \cite{donati2015} and \cite{monaco2014} found slightly super-solar abundances of Mg and Al, in agreement with the results obtained in other OCs of similar metallicity \citep{bragaglia2008, sestito2008, pancino2010, carrera2011, yong2012}.
\cite{caffau2014} estimated A(S) in only one member star of Tr\,5. According to these latter authors, this star is under-abundant in S with respect to the Galactic disk. In particular, \cite{caffau2014} claim a similar S behavior for Tr\,5 and the GC Terzan 7, which belongs to the Sagittarius dwarf galaxy.

In this work, we present A(S) for the same target analyzed by \cite{caffau2014} and another five stars in order to investigate the low S content of Tr\,5.

\subsection{Trumpler\,20}
The OC Trumpler\,20 (Tr\,20) is located in the inner disc of the MW (RA, DEC)=(12$^h$39$^m$34$^s$, $-60^\circ37'00''$) at $\sim$3.3 kpc from the Sun and $\sim$7 kpc from the Galactic center \citep{donati2014}.
\cite{platais2008} obtained (m-M)$_V$= 12.00 and E(B-V)=0.46 for Tr\,20.
The CMD for Tr\,20 shows a broadened main sequence turn off and a prominent and extended red clump (RC; \citealt{platais2012, donati2014}).
Considering the age of Tr\,20 (1.5$^{+0.2}_{-0.1}$ Gyr, \citealt{carraro2014}), these features are not easily explained by classical evolutionary models, making this object an interesting target with which to investigate the formation and the evolution of the MW disk.
According to the works in the literature, the metallicity of Tr\,20 is slightly super-solar: [Fe/H]=0.10$\pm$0.08 \citep{carraro2014, donati2014, tautvaisien2015}. 
Moreover, \cite{carraro2014} found solar abundances of $\alpha$-elements in Tr\,20, which follow the trend of giant stars located in the inner disk and other old OCs with similar metallicity.
Similarly, \cite{duffau2017} found a solar [S/Fe] ratio in Tr\,20.
We considered this OC in our work as a reference object.
\\

\section{Observational data}
\begin{figure*}
  \centering
        \includegraphics[trim= 3cm 1.5cm 3cm 2cm, clip, width=0.9\textwidth]{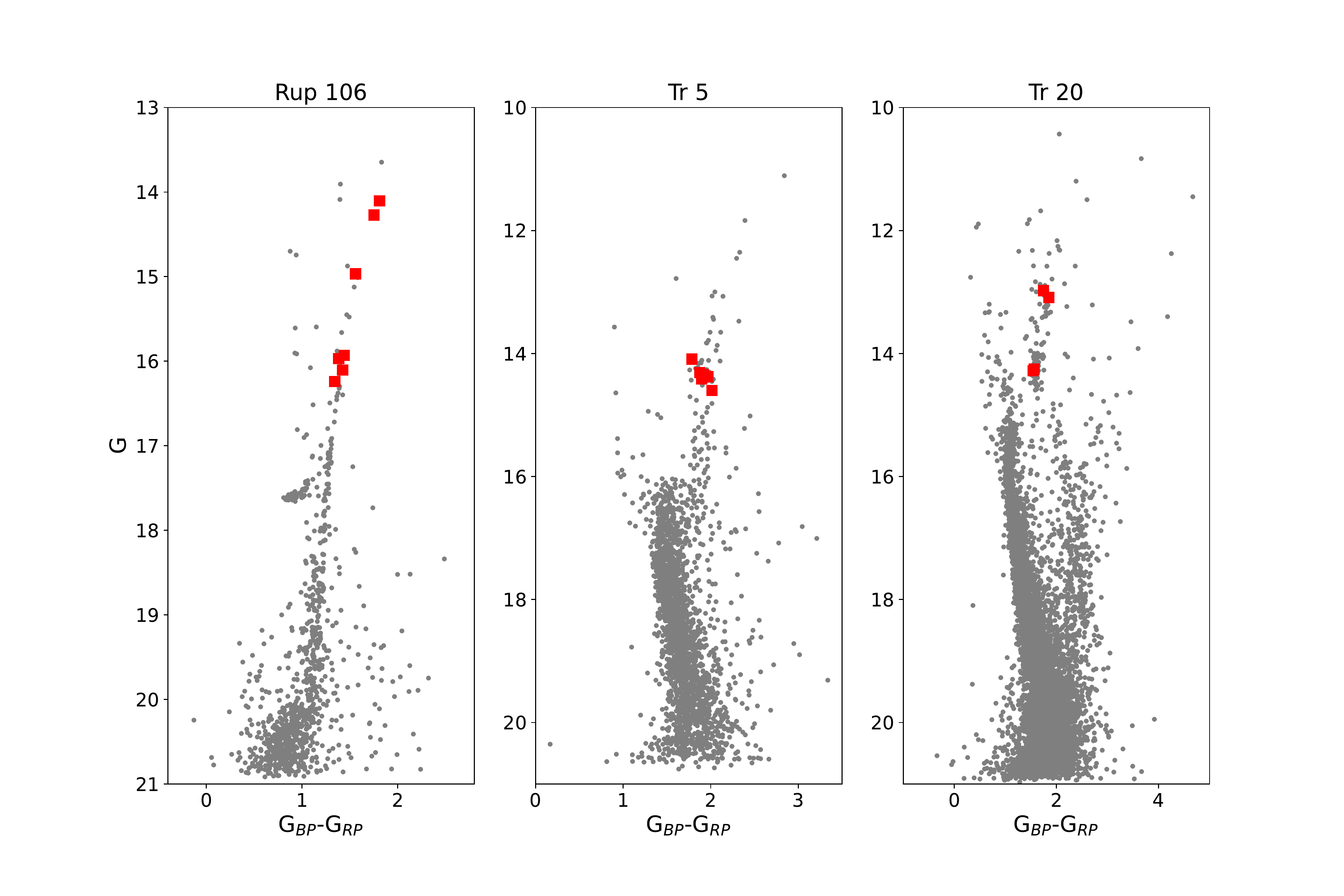}
        \caption{$G$ versus $G_{BP}-G_{RP}$ CMD of the clusters considered in this work: Rup\,106 (left), Tr\,5 (middle), and Tr\,20 (right). The sample of stars analyzed in this work is shown by the red squares.}
        \label{cmd}     
\end{figure*}

The  aim of this work is to provide the first A(S) in Rup\,106 and to investigate the behavior of S in Tr\,5. We analyzed seven RGB stars of Rup\,106, six RC stars of Tr\,5, and four giant stars (RGB and RC) of Tr\,20. Our sample of stars is shown by the red squares in the Gaia \citep{2016A&A...595A...1G} EDR3 \citep{2021A&A...649A...1G}   
$G$ versus $G_{BP}-G_{RP}$ CMDs of Figure \ref{cmd}. The ID, coordinates, and $G$,  $G_{BP}-G_{RP}$ magnitudes
of the targets are reported in Table \ref{data}.

All the data considered in this work are available in the ESO 
archive\footnote{\url{http://archive.eso.org/cms/eso-data/instrument-specific-query-forms.html}}. 
We retrieved the red-arm UVES \citep{2000SPIE.4008..534D} spectra of the targets collected with the multi-object fiber-fed FLAMES facility \citep{2002Msngr.110....1P} mounted on the ESO-VLT/UT2 telescope at the Paranal observatory (Chile). We selected the spectra observed with the 860\,nm setting (hereafter R860), which are characterized by R$\sim$47000 and cover the wavelength range where S lines lie (675-1050 nm).
These data were reduced using the ESO--CPL-based FLAMES/UVES pipeline.
We also collected the data for our targets observed with the 580\,nm setting (hereafter R580), which cover the 480-680 nm wavelength range.
In particular, we used the R580 UVES-slit data of Rup\,106 studied by \cite{villanova2013}. 
Instead, the R580 FLAMES/UVES data of Tr\,5 are those analyzed by \cite{monaco2014}. 
We also considered the spectrum of Tr 5 star \#3416 analyzed in Monaco
et al. (2014). This was obtained with the MIKE spectrograph
\citep{2003SPIE.4841.1694B} at the MAGELLAN telescope at  the Las Campanas observatory (Chile). We refer the reader to \cite{monaco2014} for details of the data reduction of this spectrum.
Finally, the R580 FLAMES/UVES data for Tr\,20 were obtained from the \textit{Gaia} ESO Survey\footnote{\url{https://www.gaia-eso.eu}} \citep{gilmore2012, randich2013}.

The dates of observation and the signal-to-noise ratios (S/Ns) of the spectra are reported in Table \ref{data}. In particular, the dates reported are indicative, as the data were collected during different nights and epochs, as specified in the notes of the table.

\section{Analysis}

\subsection{Atmospheric parameters}
In order to derive the chemical abundances from the spectra, we first need to
obtain an initial estimate of the atmospheric parameters. For this purpose, we used
the Gaia EDR3 photometry \citep{2021A&A...649A...3R} combined with the 2MASS photometry \citep{2003yCat.2246....0C}. 
This is a homogeneous and wide base photometry of the targets, which is ideal to better estimate the reddening.
Only  the stars with Gaia EDR3 proper motions compatible with cluster membership were taken into account.
We first apply the isochrone fitting method, where an isochrone is adjusted to the CMD. We used Padova isochrones \citep{Bressan2012} for this purpose.
For Ruprecht\,106, we used an age of 12 Gyr \citep{VandenBerg2013} and isochrones spanning a range of metallicities between --2.00 and --1.00 in steps of 0.05 dex.
For Trumpler 5 and Trumpler 20 instead, data were deep enough to estimate ages
from the turn-off (TO) magnitude, and so we treated this as a free parameter to be estimated during the isochrone fitting. For the two clusters, we used a metallicity range from --0.5 to +0.20, with a step of 0.05 dex.
The interstellar absorption was taken into consideration using the \cite{Cardelli1989} relation, applying a point-by-point correction to the isochrones.
In this case, the free parameters are the V band absorption (A$_V$) and the reddening-law coefficient (R$_V$).

For Ruprecht\,106, we obtained the best fit of the RGB for a distance of 22080 pc (intrinsic distance modulus of (m-M)$_0$=16.72), A$_V$=0.71, R$_V$=2.70 (which translates into E(B-V)=0.26), and a global metallicity [M/H]=-1.85.
\cite{Harris2010} gives a distance modulus in the V band of 17.25 and
a reddening E(B-V)=0.20. Our reddening and distance modulus are slightly
higher, because (m-M)$_V$=17.25 translates into  (m-M)$_0$=16.63 if E(B-V)=0.20
and R$_V$=3.1 are assumed. The global metallicity [M/Fe] we find is lower
than the iron content of the cluster ([Fe/H]=-1.68) as given by \cite{Harris2010}. This is due to the fact that [M/H] is affected not only by
the iron content, but also by the amount of all the other elements in
the atmosphere, including C, N, and O, which are the most abundant.
Our finding could mean that Ruprecht\,106 is particularly low in one of
those elements. 
Indeed, several works in the literature show that Rup\,106 is poor in Na and not enhanced in $\alpha$-elements \citep{buonanno1990, brown1997, villanova2013, frelijj2021}.

For Trumpler 5 and Trumpler 20 instead, we obtained the best fit  
for distances of 2900 and 2800 pc (intrinsic distance modulus of (m-M)$_0=$ 12.31 and 12.24, respectively), A$_V$=2.05 and A$_V$=1.55, R$_V$=3.1 for both clusters (that translates into
E(B-V)=0.66 and E(B-V)=0.50), a global metallicity of [M/H]=--0.40 and
[M/H]=+0.15, and ages of 4.0 Gyr and 1.6 Gyr, respectively.
The WEBDA database \footnote{\url{https://webda.physics.muni.cz/navigation.html}} gives
a distance of 3000 pc, a reddening of 0.58 and an age of 4.0 Gyr for
Trumpler 5, while for Trumpler 20 it gives a distance of  2420 pc, a reddening
of 0.26, and an age of 0.15 Gyr. In the case of Trumpler 5, these values are in very good  agreement with our findings, while for Trumpler 20, reddening and especially age are significantly discrepant.
We attribute this to the fact that this cluster is strongly affected by field contamination, and previous studies did not have proper motions in order to remove it. Instead, with proper motions, we were able to obtain a clean CMD and therefore to determine a reliable age.

After that, for Ruprecht\,106 we derived \teff\,$-K$ and \logg\,$-K$ relations
using the RGB part of the isochrone. We obtained temperature and  gravity for
our targets by simply entering their $K$ magnitudes into these relations.
Indeed, as all the targets of Rup\,106 are RGB stars, it is possible to define a unique relation between \teff\, and magnitude. 
Microturbulence (v$_t$)  instead was calculated using the equation
from  \cite{2016A&A...585A..75D}, which takes both temperature and gravity into account.
In particular, these authors presented a relation based on 3D atmospheric models for a sample of subgiant and dwarf stars.

For Trumpler 5 and Trumpler 20 instead, we preferred to derive temperature from the \teff\,$-(G_{BP}-K$) relation obtained from the RGB part of the isochrone. 
This is because the targets of Tr\,5 are RC, while the sample of stars in Tr\,20 includes RGB and RC.
Therefore, due to the fact that we have stars with different \teff\, at the same magnitude, it was not possible to derive a unique \teff\,--magnitude relation.
The gravity was again obtained from the \logg$-K$ relation. Also in this case, the microturbulence was calculated using the equation from  \cite{2016A&A...585A..75D}.

The obtained atmospherical parameters are listed in Table \ref{parameters}.
The typical errors in \teff\, and \logg\, are 50 K and 0.1 dex, respectively. These values were derived introducing the photometrical uncertainties in the \teff\,$-K$, \teff\,$-(G_{BP}-K$) and \logg$-K$ relations. We stress the fact that these are internal uncertainties; external or systematic errors can be larger and are mainly related to uncertainties on isochrone fitting.

\subsection{Radial velocities}
As the spectra of the targets were collected during different nights or epochs, we computed the radial velocities (RVs) of each spectrum.
We used the IRAF/\textit{fxcor}\footnote{IRAF is distributed by the National Optical Astronomy Observatories, which are operated by the Association of Universities for Research in Astronomy, Inc., under cooperative agreement with the National Science Foundation.} task to cross correlate the observed spectra with a synthetic one. 
The theoretical spectra were computed with the code \textsc{SYNTHE} \citep{kurucz1993b,kurucz2005} using ATLAS9 model atmospheres \citep{kurucz1993} based on opacity distribution functions (ODFs) by \cite{castelli2003}, with the atmospheric parameters in Table \ref{parameters}.

Our choice to use ATLAS models is driven by the fact that the code is 
open source and we can compute models for the desired parameters. 
We prefer to compute models, of which we can check the convergence, 
rather than to use pre-computed grids of models, which provide no
information on the convergence. Furthermore, to use a model grid it
is necessary to interpolate and it is not guaranteed that the 
interpolated model satisfies the constraint of constant flux and zero-flux derivative
from layer to layer.
The referee raised the issue of the pertinence of using plane parallel
models versus spherical models for our stars. 
The effect is very small. We compared synthetic spectra
of the S\,I Mult.1 lines,
computed  with {\tt turbospectrum}
\citep{2012ascl.soft05004P} (which can treat both spherical and plane parallel
radiative transfer)
for a MARCS \citep{marcs} spherical model
of \teff\ 4250 K, log g = 0.5, microturbulence 2\,\kms ,
metallicity --1.5, and $\alpha$ enhancement +0.4, and an ATLAS 9 model with
identical parameters, which were computed by us.
These parameters
are close to those of stars \#\,1614 and \#\,2004
in Ruprecht 106, which are the two most luminous giants
in our sample, for which the sphericity effects are largest. 
In terms of abundance, the difference
between the two synthetic spectra can be quantified as 0.08\,dex,
in the sense that use of plane parallel models results in smaller
abundances. We underline that the for plane parallel models,
ATLAS and MARCS models are well known to be equivalent
\citep[e.g.,][Appendix A]{Bonifacio09}.
It is impossible to compare a MARCS spherical with a MARCS plane parallel
model with the same parameters using the models available online\footnote{\url{https://marcs.astro.uu.se/}},
because there is no overlap of the two classes of models.
Nevertheless, we believe our test provides the correct order of magnitude of the sphericity effect.
This is an upper limit  for our sample of stars, as all have higher
surface gravities. 
It has not been demonstrated that spherical 1D models provide
a better modeling of the optical and near-infrared (NIR) spectra of K giants. 
We conclude that there is no reason to use
spherical 1D models for our analysis.
 We decided to use ATLAS 9 models rather than ATLAS 12 models (which use opacity sampling
instead of ODFs) since they can be computed faster (a factor of 30) and
for this kind of stars the two class of models are equivalent, as discussed in Appendix \ref{app_9_12}.

We corrected to the rest frame and combined the spectra of each star.
The RVs obtained in this work for the sample of stars are listed in Table \ref{RV}. In particular, the RVs reported are the mean value obtained from each exposure. In the case of Tr\,5 R580 data, we used and reported the RVs estimated by \cite{monaco2014}.

\subsection{Iron abundances}
To estimate the metallicity of the targets, we created our iron (Fe I and Fe II) linelist. 
For this purpose, we used the R580 datasets to create an iron linelist for each cluster by comparing observational and synthetic spectra. The linelists of the three clusters were joined and matched with the Gaia-ESO  linelist \citep{GAIAlinelist} to get more precise atomic data. The final linelist is composed of 258 Fe I and 42 Fe II lines, for a total of 300 features in the wavelength range 480-680 nm.

The iron abundances of the targets were derived from line equivalent widths (EWs) using the code \textsc{GALA} \citep{mucciarelliGALA}. We measured the EW with the software \textsc{DAOSPEC} \citep{daospec}. We want to emphasize that each line was carefully checked against blending before being selected for the estimation of iron abundances. Moreover, we selected Fe lines with EW $< 100  \rm{m\AA}$ and excitation potential $\chi > 1.2 $ eV. 
We investigated the uncertainty on A(Fe) due to atmospheric parameter errors.
From the process adopted to estimate the atmospheric parameters, the typical uncertainties are $\sigma_T \sim 50$ K, $\sigma_{\logg} \sim 0.1 $ and $\sigma_{\xi} \sim 0.1$ km s$^{-1}$.
We found that the A(Fe) uncertainties due to $\sigma_T$, $\sigma_{\logg}$, and $\sigma_{\xi}$ are 0.05, 0.02, and 0.02 dex, respectively. A variation about 0.1 dex in metallicity leads to an A(Fe) uncertainty of 0.01 dex.
Table \ref{parameters}  reports the stellar metallicities obtained; we assume the solar value A(Fe)$_{\odot}=7.52\pm0.06$ \citep{Caffau2011} based on the investigations by \cite{caffau2007a} and \cite{caffau2007b}. 

The dataset for Tr\,5 does not include the R580 spectra of all the stars. Furthermore, we were not able to measure the [Fe/H] ratio from these data for $\#$3678 due to the low S/N of the spectrum. For these reasons, we created an Fe linelist and estimated the metallicity of Tr\,5 targets using the data collected with the R860 setting. In the case of the star $\#$3416, we used the MIKE spectrum due to its higher S/N. For $\#$1318, the [Fe/H] values obtained from R580 and R860 datasets are comparable within the errors. Similarly, for  $\#$3416, we found good agreement between the metallicities estimated from MIKE and R860 data.
This implies that there is not a systematic offset between [Fe/H] as derived from different datasets. This supports the reliability of the metallicity estimated from only the R860 data. On the other hand, the final metallicities of the stars $\#$1318 and $\#$3416 were measured using Fe linelists created from both datasets (R580 or MIKE, and R860).

We found a mean [Fe/H]= --1.37 $\pm$ 0.11 for Rup\,106, in agreement with previous works \citep{francois1997, brown1997, pritzl2005, villanova2013, frelijj2021}.
The mean metallicy of Tr\,5 is [Fe/H]= --0.49 $\pm$ 0.14. This result is consistent with that of \cite{cole2004, piatti2004, carrera2007, kim2009} and \cite{donati2015}.
According to several works in the literature \citep{platais2008, carraro2014, tautvaisien2015, donati2014}, Tr\,20 is characterized by a slightly super-solar metallicity. We obtained an iron content for the OC Tr\,20 of about 0.06 $\pm$ 0.15.

\subsection{Sulfur abundances}
The S I lines that we analyzed in this work are reported in Table \ref{par_ato}. 
Thanks to the high resolution and the high S/N of the R860 data, we were able to consider the weak lines of multiplet (Mult.) 8 and 6 besides those of Mult. 1. However, these lines were not identifiable in the stellar spectra of Rup\,106 due to the low metallicity of the cluster. Consequently, the A(S) in Rup\,106 were estimated from Mult. 1 features only.

The possible telluric lines contamination was taken into account using TAPAS atmospheric transmission spectra \citep{Bertaux2014}. We want to underline that, as opposed to the other clusters, Tr\,5 data were observed in different epochs. In this case, we created a TAPAS profile for each epoch. As S lines of Mult. 6 and 8 are not affected by telluric lines, we combined the spectra of each epoch to measure A(S) from these lines. The line at 9212 $\AA$ was not found to be contaminated in October and November epochs, and so these last were combined to calculate A(S). Similarly, the epochs of December were combined to estimate the A(S) from the lines at 9228 $\AA$ and 9237 $\AA$. 
In the case of Rup\,106 and Tr\,20, the Mult. 1 in the different exposures of each star were not found to be contaminated by telluric lines. Consequently, we combined all the R860 data of Rup\,106 and Tr\,20 stars to estimate their A(S).

\begin{figure*}
  \centering
        \includegraphics[trim=2.5cm 0.8cm 3.2cm 2cm, clip,width=0.8\textwidth]{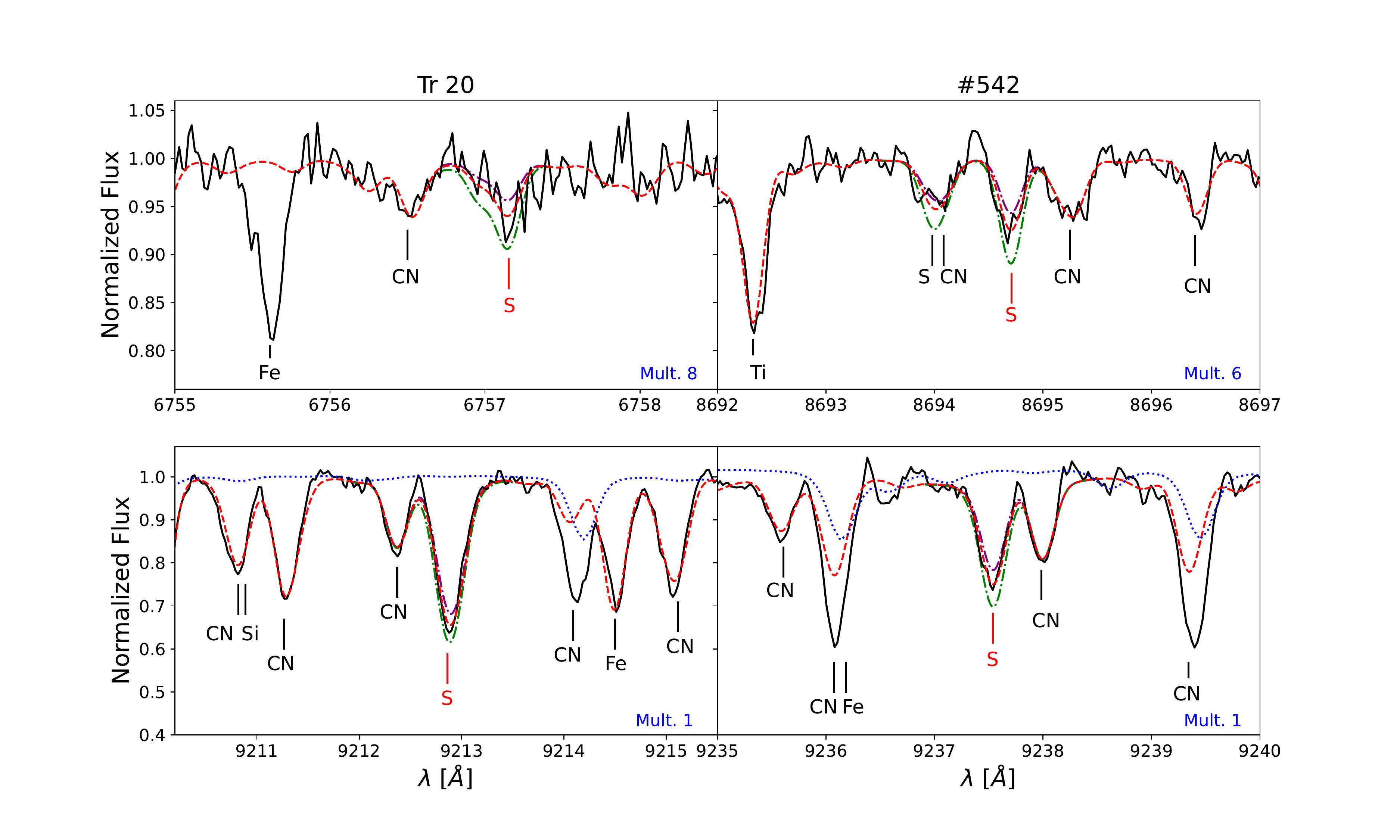}
        \caption{Observed spectrum of the Tr\,20 star $\#542$ (black) superimposed with the best-fit synthetic spectrum (dashed red). The green and purple profiles are synthetic spectrum with +0.2 dex and -0.2 dex in A(S), respectively. The TAPAS profile is also shown (dotted blue). The red vertical lines indicate the sulfur features.}
        \label{fit}     
\end{figure*}

Once the suitability of the S lines had been  evaluated, we measured A(S) by spectrosynthesis. Employing our code SALVADOR (Mucciarelli in prep.), the chemical abundance is found when the minimum $\chi^2$ between the observed and the synthetic spectra is reached.
Figure \ref{fit} shows the observed spectrum (black) of the Tr\,20 star $\#542$ with the best-fit synthetic spectrum (dashed red) and the TAPAS profile (dotted blue) superimposed.
We estimated the [S/Fe] ratio considering the solar value A(S)$_\odot=7.16$ \citep{Caffau2011}. The mean A(S)$_{\rm LTE}$ with their standard deviations of each target are reported in Table \ref{lines}.
We investigated the uncertainty in chemical abundance determination due to the  error on stellar parameters.
Adopting the typical uncertainties $\sigma_T \sim 50$ K, $\sigma_{\logg} \sim 0.1, $ and $\sigma_{\xi} \sim 0.1$ km s$^{-1}$, we found errors on A(S) of about 0.06 dex , 0.04 dex, and 0.02 dex, respectively. Finally, considering $\sigma_{\rm [Fe/H]}$= 0.1, we found $\sigma_{\rm A(S)} = 0.02$.

\subsection{NLTE corrections}
Following \cite{Takeda2005}, we derived the nonlocal thermodynamic equilibrium (NLTE) corrections for lines of Mult. 6 and 1. 
Table \ref{lines}  reports the LTE and NLTE A(S) obtained from each line of our targets and the mean A(S)$_{\rm LTE}$ and A(S)$_{\rm NLTE}$ with their standard deviations.
The mean LTE and NLTE line-to-line scatter for Rup\,106 stars are both 0.06 dex. We obtained a mean NLTE correction of $<\Delta> \sim -0.18$ and $<\Delta> \sim -0.11$ for the lines at 9212 $\AA$ and 9228 $\AA$, respectively. 
We find a mean LTE line-to-line scatter of 0.10 dex for Tr\,5.
As expected from \cite{korotin2020}, Mult. 6 is less affected by LTE deviations,$-0.05<\Delta<-0.02$, than Mult. 1, $-0.47<\Delta<-0.25$. As a consequence, the mean NLTE line-to-line scatter increases to 0.14 dex due to the difference between the A(S) values obtained from Mult. 8 and Mult. 1 lines.
This is also the case for Tr\,20 stars, where we find $<\Delta> \sim -0.05$ for Mult. 6 line, $<\Delta> \sim -0.3$ and $<\Delta> \sim-0.23$ for lines at 9212 and 9228 $\AA$, respectively.
The mean LTE and NLTE line-to-line scatter for Tr\,20 stars are respectively 0.01 dex and 0.16 dex.

\begin{figure}
  \centering
        \includegraphics[trim= 2cm 2cm 3.5cm 2cm, clip, width=0.5\textwidth]{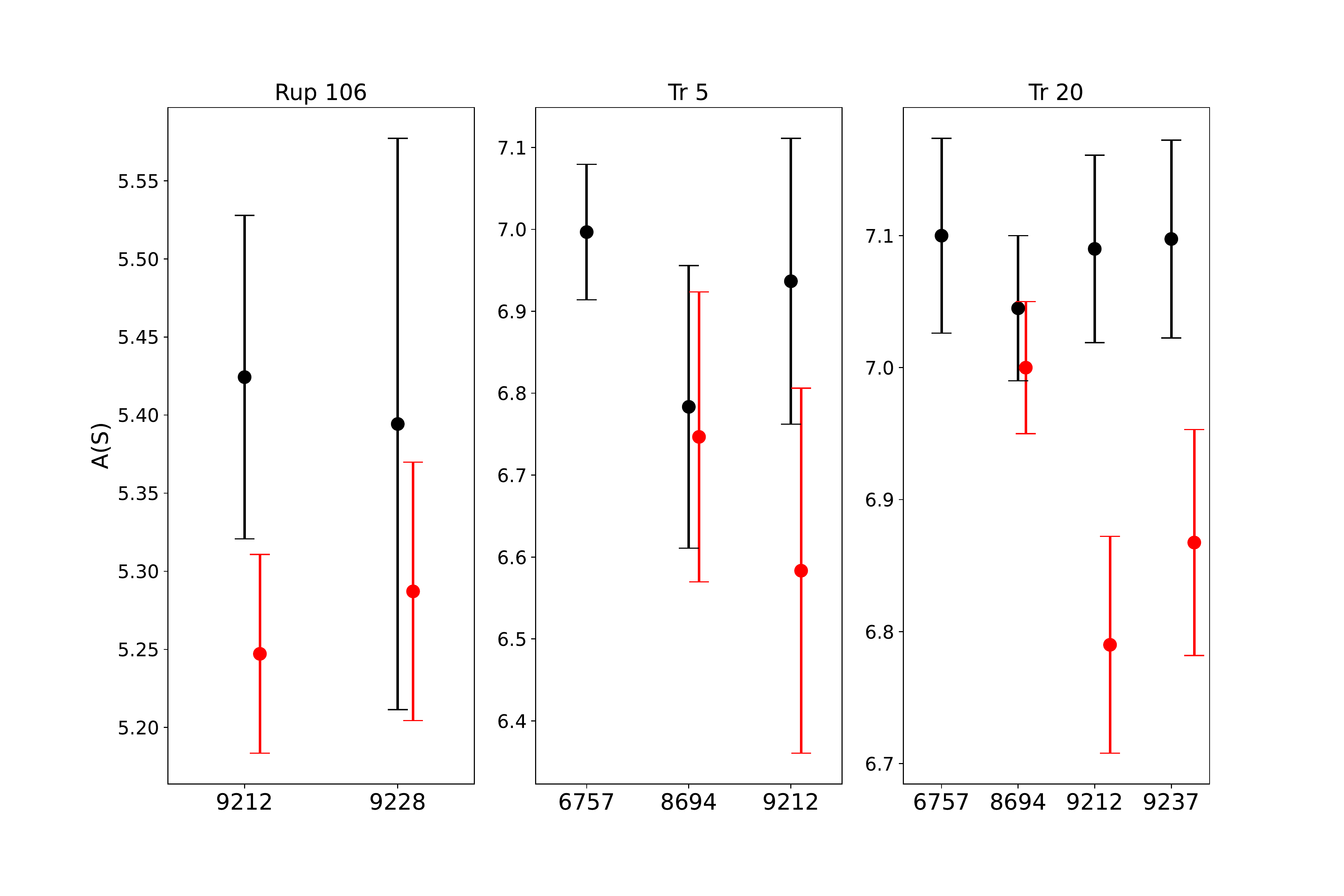}
        \caption{Mean LTE (black) and NLTE (red) sulfur abundances obtained from the different lines in the spectra of Rup\,106 (left panel), Tr\,5 (middle panel), and Tr\,20 (right panel). The error bars are the standard deviation of the A(S) measures for each line, from all the stars in a cluster.}
        \label{comp_lines}      
\end{figure}

In order to better see the effect of NLTE corrections, we compared the mean LTE (black) and NLTE (red) A(S) obtained for the stars of each cluster from the  different lines in Figure \ref{comp_lines}. The error bars are the standard deviation of the A(S) obtained from the lines.
It is clear from this plot that the A(S)$_{\rm LTE}$ obtained from the different Mult. are consistent between each other. In contrast, this agreement is no longer found when comparing A(S)$_{\rm NLTE}$ obtained from Mult. 8, 6, and 1. 
In conclusion, it is  evident that the increased scatter in A(S)$_{\rm NLTE}$ is due to NLTE corrections.

This may lead us to question the NLTE corrections; specifically, we suspect the predicted NLTE effects for Mult. 1 lines in the Tr\,5 and Tr\,20 stars are overestimated.
We suggest the problem lies in collisions with H atoms, which is treated by
\citet{Takeda2005} using the Drawin formalism \citep{drawin1969} as generalized
by \citet{steenbock1984}.
Although the Drawin formalism is usually employed, in the absence of theoretical
or experimental data on the cross-sections of collisions with H atoms, 
\citet{barklem2011} showed that such formalism does not contain the essential physics to explain H atom collisions from a quantum mechanical point of view.
\cite{Takeda2005} used a scaling factor $h$, which corresponds to the logarithm of the correction factor applied to the results of the Drawin formalism.
\cite{Takeda2005} analyzed the cases of $h = +1, 0, -1, -2,$ and $-3$, where $h=0$ corresponds to the classical formula. 
The NLTE effect increases as $h$ decreases. Increasing the number of collisions drives the system toward LTE. Moreover, NLTE corrections are negative and larger (in absolute value) 
for low \logg, high \teff, and low metallicity. 
\cite{Takeda2005} tabulated NLTE abundance corrections ($\Delta_h$) for lines of Mult. 6 and 1 for an extensive grid of 210 model atmospheres ($\teff\,=4000-7000 K, \logg=1.0-5.0, \xi=2 \,km s^{-1}, $[Fe/H]$= +0.5, 0.0, -0.5, -1.0, -2.0, -3.0, -4.0$).
From Table 2 of \cite{Takeda2005}, the corrections $\Delta_{+1}=-0.14$, $\Delta_0=-0.32$, $\Delta_{-1}=-0.46$, $\Delta_{-2}=-0.52$, $\Delta_{-3}=-0.54$ correspond to the model with \teff \,=4500, \logg
=2.0, [Fe/H]=0.0 for the S line at 9212 $\AA$.
In Figure \ref{comp_lines}, it is possible to see that the $<\Delta>$ for 9212 $\AA$ line is around 0.3 dex, in agreement with what is expected using $h=0$. However, this kind of correction leads to a discrepancy between the A(S) estimated from lines at 6757 $\AA$ and 9212 $\AA$. Nevertheless, this difference is reduced when assuming $h=+1$, and a lower (negative) NLTE correction for 9212 $\AA$ line is obtained.
We conclude that the disagreement in A(S)$_{\rm NLTE}$ obtained from Mult. 8, 6, and 1 (shown in Fig. \ref{comp_lines}) might be related to the role of collisions with H atoms and the inadequacy of the Drawins formalism. 
The close agreement of the LTE abundances derived from the 
different multiplets suggests that they are all formed close to LTE conditions. 

\subsection{Comparison with the literature}
\begin{figure}
  \centering
        \includegraphics[trim= 1cm 1.6cm 2.2cm 2.4cm, clip, width=0.5\textwidth]{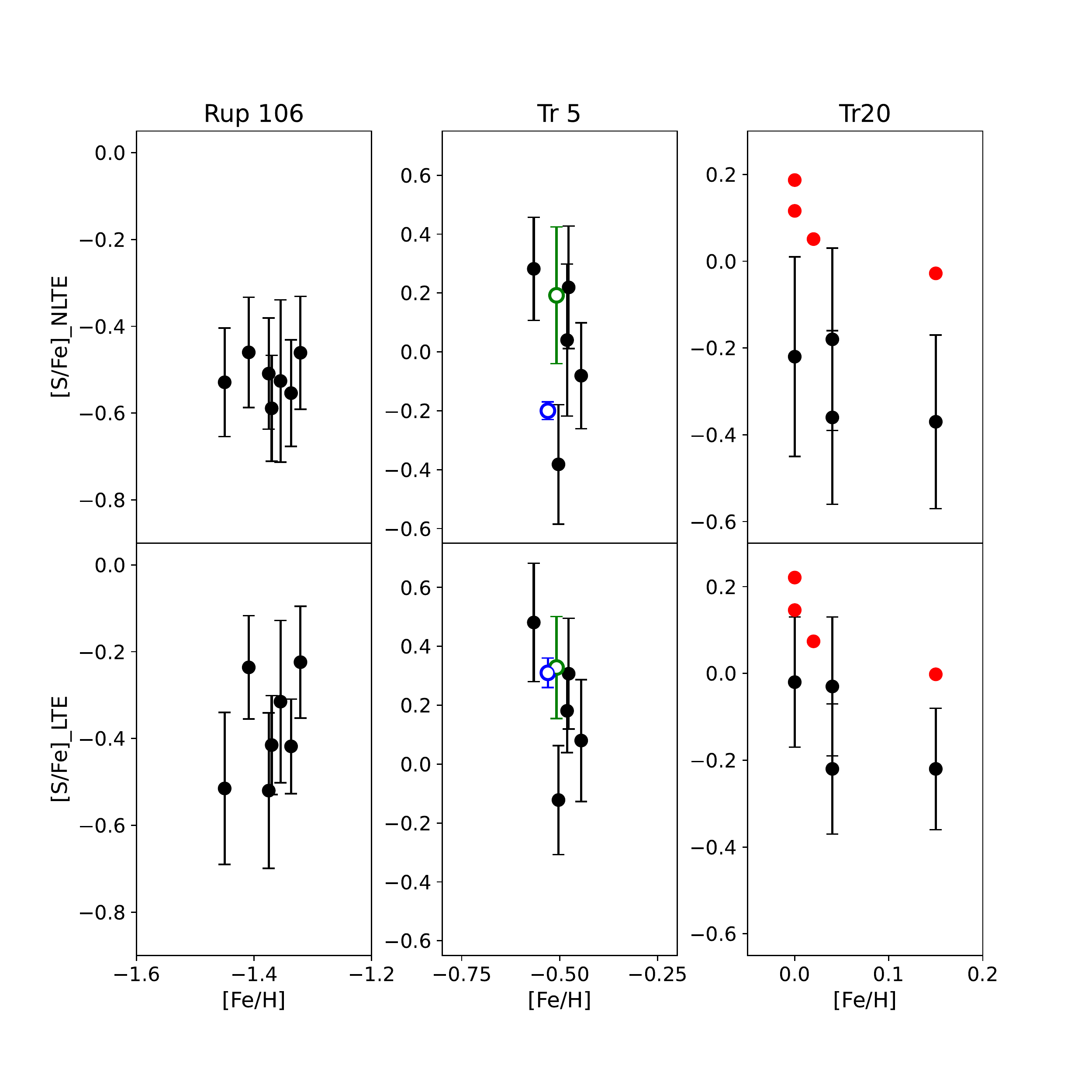}
        \caption{[S/Fe] versus [Fe/H] diagram before (bottom panels) and after (top panels) NLTE corrections applied to the A(S) obtained in this work for Rup\,106 (left), Tr\,5 (middle), and Tr\,20 (right) stars. In the middle panels, the results obtained by \cite{caffau2014} for the star $\#$3416 are reported in blue. Our results for the same star are highlighted in green. In the right panels, our results are compared with those of Duffau et al. (\citeyear{duffau2017}, red points). }
        \label{results} 
\end{figure}

Figure \ref{results} shows the [S/Fe] versus [Fe/H] diagrams of the targets before (bottom panel) and after (top panel) the NLTE corrections. 
The error bars were calculated with the error propagation, combining the uncertainties on A(S) and [Fe/H]  in quadrature.

We find Rup\,106 to be the object with the lowest S content in both diagrams. 
These results represent the first A(S) measurement for the stars of Rup\,106. Therefore, a comparison with the literature cannot be shown.

\cite{caffau2014} analyzed the star $\#$3416 of Tr\,5 (blue circle in Figure \ref{results}).
In this work, we analyzed the same star (green circle in Figure \ref{results}) and a further five targets in Tr\,5. Our [S/Fe]$_{\rm LTE}$ value for star $\#$3416 is in good agreement with the findings of \cite{caffau2014}. On the other hand, after NLTE corrections, our value of [S/Fe]$_{\rm NLTE} = 0.19$ is higher than that of \cite{caffau2014} ([S/Fe]$_{\rm NLTE} = -0.20$) by $\sim$0.4 dex. We investigated the origin of this discrepancy.
The first factor to take into account is the features used.
For $\#3416,$ we measured A(S) from lines at 6757  $\AA$ (Mult. 8), 8694 $\AA$ (Mult. 6), and 9212 $\AA$ (Mult. 1), while \cite{caffau2014} used the two Mult. 1 lines at 9212 $\AA$ and 9228 $\AA$.
As previously shown, the line of Mult. 6 is less affected by LTE deviation than those of Mult. 1. Instead, the NLTE corrections for the line at 6757 $\AA$ are negligible \citep{Takeda2005}. 
Consequentially, our result for A(S)$_{\rm NLTE}$ is given considering two lines almost at LTE and one line that is strongly affected by LTE deviations. On the other hand, \cite{caffau2014} used only lines with high NLTE corrections.
For  $\#3416$, we obtain A(S)$_{\rm LTE}=$ 7.05, 6.85, and 7.04 from Mult. 8, 6, and 1, respectively. The NLTE corrections lead to A(S)$_{\rm NLTE}=$ 6.81 and 6.67 from Mult. 6 and 1, respectively. Thus, the main contribution to our NLTE correction for star $\#$3416 is given by the $\Delta =-$0.37 for line at 9212 $\AA$. 
Instead, \cite{caffau2014} found $\Delta \sim -0.51$ for Mult. 1 lines.

The atmospheric parameters is another factor to consider. We analyzed the effect of varying the  atmospheric
parameters on NLTE corrections.
Indeed, we found \teff=4869, \logg=2.52, $\xi$=1.27, and [Fe/H]=-0.51 for $\#3416$, while \cite{caffau2014} adopted \teff=4870, \logg=2.05, $\xi$=1.33, and [Fe/H]=-0.53.
Therefore, it is the surface gravity that is different by almost 0.5 dex.
According to \cite{Takeda2005} and \cite{korotin2020}, the NLTE corrections and the surface gravity are inversely proportional. Varying the surface gravity of 0.5 dex, we obtained a difference in the A(S)$_{\rm NLTE}$ from the lines of $\sim$0.18 dex. We infer that the difference between the \cite{caffau2014} [S/Fe]$_{\rm NLTE}$ values and ours is due to the different features used and to the different surface gravity used.

In the right panels of Figure \ref{results}, we compared our results for Tr\,20 and those obtained by \cite{duffau2017}. In general, we find a lower [S/Fe] value by $\sim$0.19 dex.
Also in this case, we investigated the origin of this difference.
\cite{duffau2017} used Mult. 8 lines, which are almost unaffected by LTE deviations.
Instead, our results were obtained from Mult. 8, 6, and 1. Table \ref{lines} shows the A(S)$_{\rm LTE}$ and A(S)$_{\rm NLTE}$ obtained from each line.
As mentioned above, we found $<\Delta> \sim -0.05$ for Mult. 6 lines. On the other hand, the NLTE corrections of Mult. 1 lines lead to a difference of $\sim 0.2$ dex between S abundances obtained from them and Mult. 8 lines. 
Considering the atmospheric parameters, the mean difference between our values and those obtained by \cite{duffau2017} are $<\Delta \teff>= 97K, <\Delta \logg>=0.05, <\Delta \xi>=0.29 \, \rm km s^{-1}, <\Delta [Fe/H]>=0.05$.
According to our uncertainty analysis, these values lead to a difference in S abundance of about 0.18 dex.
In conclusion, the different features and the atmospheric parameters adopted explain the difference between our analysis and that of \cite{duffau2017}.

\section{Discussion}
So far, only a few clusters of the MW were considered to investigate the behavior of S. In this work, we present S abundances in the GC Rup\,106 and the OCs Tr\,5 and Tr\,20.

We decided to compare our results (red squares) with the literature in the [S/Fe]$_{\rm NLTE}$ versus [Fe/H] (Figure \ref{sfe_nlte}) and [S/Fe]$_{\rm LTE}$ versus [Fe/H] (Figure \ref{sfe_lte}) diagrams. These figures are updated versions of Fig. 9 of \cite{duffau2017}. 
The blue open circles are the results obtained by \cite{jonsson2011} for metal-poor giant stars using Mult. 3 and the forbidden [SI] line at 1082 nm. 
The results obtained by \cite{skuladottir2015} for the stars of the Sculptor dwarf spheroidal (dSph) galaxy are shown as blue stars. These latter authors estimated A(S) from Mult. 1 features.
The gray stars are Galactic bulge stars analyzed by \cite{lucertini2022} using Mult. 8, 6, and 1.
\cite{ecuvillon2004} estimated S abundances in a large set of planet host stars and solar-type dwarfs with no known planetary-mass companions. Their S abundances, which were obtained from Mult. 8, are reported in Figure \ref{sfe_nlte} as green squares.
The black diamonds symbols represent the Galactic halo GCs M4, M22, and M30 analyzed by \cite{kacharov2015} using Mult. 3 lines.
The average in metallicity and [S/Fe] for the stars with [Fe/H]$<-2.9$ found by \cite{spite2011} from Mult. 1 lines is reported as "LP stars" (where LP  refers to the ESO large programme "First Stars" described in \citealt{cayrel2004}).
\cite{matrozis2013} analyzed metal-poor giant stars in the metallicity range $-2.5<$[Fe/H]$<0.0$ in order to obtain A(S) from the forbidden [SI] line at 1082 nm. The MP star point represents the average in metallicity and [S/Fe] for the stars with [Fe/H]$<-1.0$ from this work.
The black points are the GCs and OCs results in the literature, which were obtained using Mult. 1 and 8: NGC\,6397 from \cite{koch2011}; NGC\,6752 and 47\,Tuc from \cite{sbordone2009}; Terzan\,7 from \cite{caffau2005b}; and M4, Trumpler\,5, NGC\,5822, and NGC\,2477 from \cite{caffau2014}. 
The [S/Fe] values measured by \cite{duffau2017} are displayed in Figure \ref{sfe_nlte} as crossed black squares. 
Finally, we updated the diagram adding the results from the \textit{Gaia-}ESO Survey Data Release 5 (GES DR5, black triangles). In particular, we selected targets with a membership probability of MEM3D $>0.9$ and reliable S abundances ($\sigma_{A(S)}\leq 0.2$ dex).
In order to facilitate the comparison with our outcomes, we highlighted the results for Tr\,5 from \cite{caffau2014} (red point) and those for Tr\,20 from \cite{duffau2017} (crossed red square) and GES DR5 (red triangle).

\subsection{{\rm [S/Fe]$_{\rm NLTE}$} versus {\rm [Fe/H]} diagram}
\begin{figure*}
  \centering
        \includegraphics[trim= 2cm 1.5cm 3cm 3cm, clip, width=0.6\textwidth]{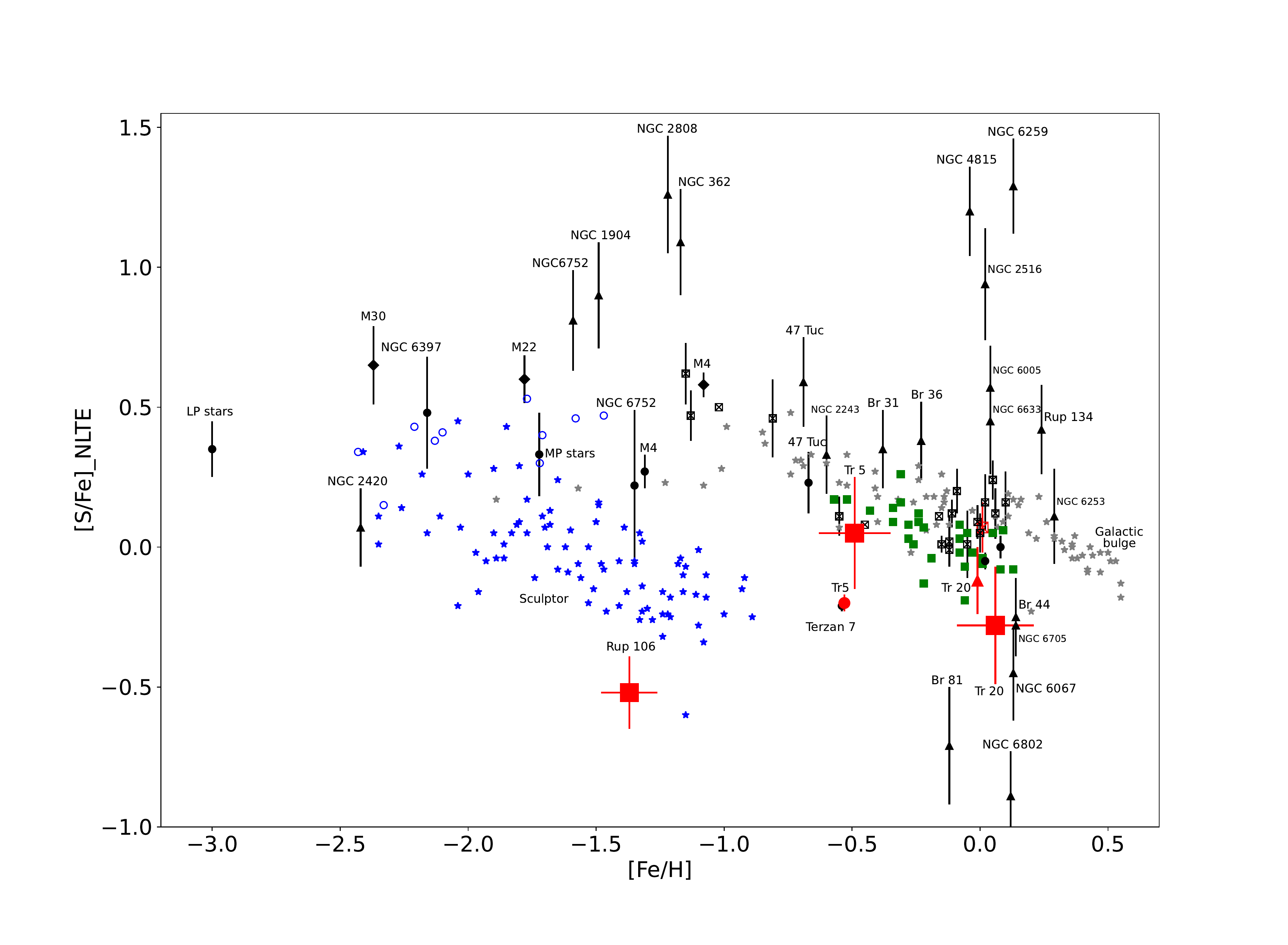}
        \caption{[S/Fe]$_{\rm NLTE}$ vs. [Fe/H] for the clusters analyzed in this work (red squares) compared to the literature. The blue open circles are  the results obtained by \cite{jonsson2011} for metal-poor stars. The Sculptor dSph galaxy \citep{skuladottir2015} and the Galactic bulge \citep{lucertini2022} are shown as blue and gray stars, respectively. The green squares are the results of \cite{ecuvillon2004} for planet-host stars  and solar-type dwarfs. The black diamond symbols represent the Galactic halo GCs M4, M22, and M30 analyzed by \cite{kacharov2015}. The extremely metal-poor stars from the large program "First stars" \citep{cayrel2004} analyzed by \cite{spite2011} are shown by the LP point. The point MP stars are targets with [Fe/H]<-1.0 analyzed by \cite{matrozis2013}. The black points are GCs and OCs: NGC\,6397 \citep{koch2011}; NGC\,6752 and 47\,Tuc \citep{sbordone2009}; Terzan 7 \citep{caffau2005b}; M\,4, Trumpler\,5, NGC\,5822, and NGC\,2477 \citep{caffau2014}. The clusters from \cite{duffau2017} and GES DR5 are shown as crossed squares and triangles, respectively. All the points have been scaled to our adopted solar abundances A(S)$_\odot$ = 7.16 and A(Fe)$_\odot$ = 7.52 \citep{Caffau2011}.}
        \label{sfe_nlte}        
\end{figure*}

Tr\,20 is characterized by nearly solar [$\alpha$/Fe] ratios \citep{magrini2014, carraro2014, donati2014}.
Among the different works in the literature, \cite{duffau2017} and GES DR5 are the only ones that estimated S abundances in Tr\,20.
These works confirmed that S behaves like the other $\alpha$-elements in this OC.
We find an S content of Tr\,20 of about [S/Fe]$_{\rm NLTE}$= --0.28$\pm$0.21.
In Figure \ref{sfe_nlte}, this value (red square) is in agreement within errors with that from GES DR5 (red triangle). On the other hand, it is lower than that obtained by \cite{duffau2017} (crossed red square) by $\sim$ 0.19 dex. We claim that this discrepancy is due to the choice of atmospheric parameters and lines analyzed.

Figure \ref{sfe_nlte} reveals the low S content of Rup\,106 (red square) with respect to the other Galactic objects of similar metallicity. In particular, we obtained a mean [S/Fe]$_{\rm NLTE}$= --0.52$\pm$0.13 for Rup\,106.
This confirms that S behaves like the other $\alpha$-elements in this GC.
Indeed, the over-deficiency of $\alpha$-elements in Rup\,106 was already assessed by \cite{brown1996, brown1997}, \cite{pritzl2005}, \cite{villanova2013}, \cite{francois2014},  and \cite{frelijj2021}.
The low content of $\alpha$-elements suggests that Rup\,106 experienced different nucleosynthesis processes from the majority of the Galactic halo and nearby clusters.
Therefore, it is reasonable to propose an extragalactic origin for this object, as suggested by several works in the literature \citep{lin1992, brown1996, villanova2013, francois2014, frelijj2021, callingham2022}.

For Tr\,5 stars (red square), we found a mean [S/Fe]$_{\rm NLTE}$ value of 0.05$\pm$0.20. As shown in Figure \ref{sfe_nlte}, this value is $\sim$ 0.4 dex higher than that of \cite{caffau2014} (red point).
\cite{caffau2014} concluded that the Tr\,5 star $\#$3416 is S-poor with respect to the Galactic clusters at the same metallicity. Moreover, these authors found a similar behavior for Tr\,5 and the Sagittarius cluster Terzan 7 in the [S/Fe]$_{\rm NLTE}$ versus [Fe/H] diagram. According to our results, the S content of Tr\,5 is higher than that of Terzan 7 by $\sim$0.16 dex. In Figure \ref{sfe_nlte}, Tr\,5 lies close to the Galactic OCs NGC\,2243 and Berkely\,25 analyzed by Duffau et al. (\citeyear{duffau2017}, black crossed squares). 
Comparing our result with those of \cite{lucertini2022}, Tr\,5 is located in the region of the [S/Fe]$_{\rm NLTE}$ versus [Fe/H] diagram that describes the Galactic thick disk.
In conclusion, we can confirm that Tr\,5 shows a similar behavior to that of the Galactic disk.

\subsection{{\rm [S/Fe]$_{\rm LTE}$} versus {\rm [Fe/H]}  diagram}
\begin{figure*}
  \centering
  \includegraphics[trim= 2cm 1.5cm 3cm 3cm, clip, width=0.6\textwidth]{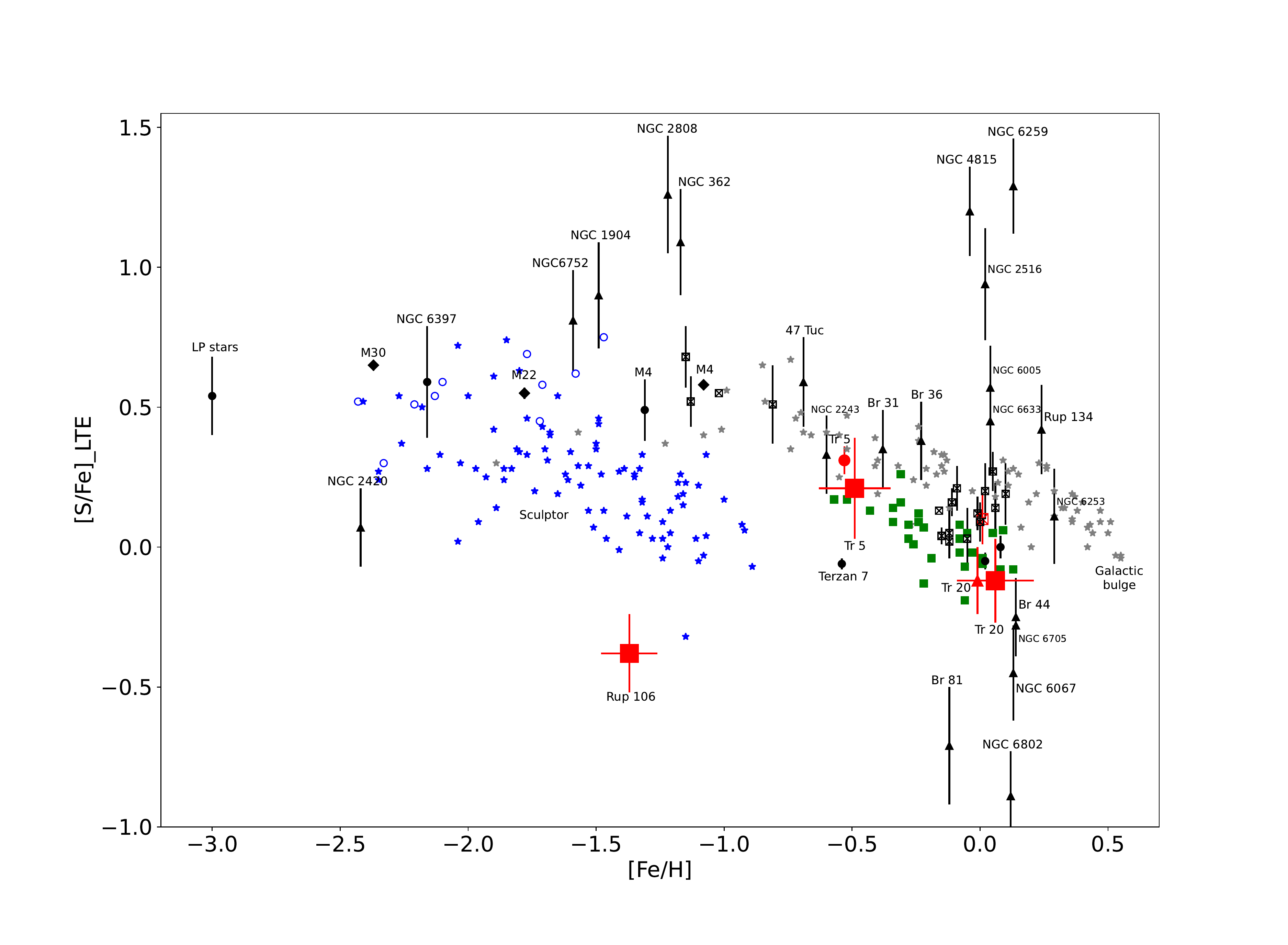}
        \caption{Same as Figure \ref{sfe_nlte} in the [S/Fe]$_{\rm LTE}$ vs. [Fe/H] diagram.}
        \label{sfe_lte} 
\end{figure*}

We reproduced Figure \ref{sfe_nlte} in the [S/Fe]$_{\rm LTE}$ versus [Fe/H] diagram (Figure \ref{sfe_lte}) with the available data. The samples displayed are the same as those described above.

In Figure \ref{sfe_lte}, Tr\,20 (red square) is in agreement within the errors with the findings of Duffau et al. (\citeyear{duffau2017}, crossed red square) and GES DR5 (red triangle).
In the [S/Fe]$_{\rm LTE}$ versus [Fe/H] diagram, Rup\,106 is characterized by a lower S content than the Galactic clusters and the Local Group Sculptor dSph galaxy (blue stars). This supports the extragalactic origin of Rup\,106. 
The Magellanic Clouds \citep{lin1992}, Sagittarius dSph galaxy \citep{bellazzini2003, frelijj2021}, and the Helmi stream \citep{massari2019, bajkova2020} were proposed as progenitors of Rup\,106.
As no S abundances were estimated in these objects, a direct comparison is impossible.
Consequently, to date, the investigation of S does not provide further clues as to the origin of Rup\,106. 

Our result (red square) and that obtained by \cite{caffau2014} (red point) for Tr\,5 are in agreement within their errors. 
In the [S/Fe]$_{\rm LTE}$ versus [Fe/H] diagram, our result for Tr\,5 lies in the Galactic thick-disk region.

\section{Summary and conclusions}
Here, we present our study of the sulfur content of the GC Rup\,106 and the OC Tr\,5.
In particular, we investigated the S abundances in Rup\,106 for the first time to investigate the low level of $\alpha$-elements in this object.
As the only star of Tr\,5 studied so far has been claimed to be S-poor, we analyzed a larger sample in order to obtain clues as to the behavior of S in Tr\,5.
This work provides the Fe and S abundances for Rup\,106, Tr\, 5, and the OC Tr\,20. This latter object was taken into account for reference, and its results are consistent within the errors with those of \cite{duffau2017} and GES DR5.
Below we summarize our main findings for the clusters analyzed in this work.

\begin{itemize}
        \item The Fe contents of Rup\,106 and Tr\,5 are --1.37$\pm$0.11 and -0.49$\pm$0.14, respectively.

        \item Rup\,106 is S-poor with respect to the other Galactic clusters of similar metallicity. We obtain [S/Fe]$_{\rm NLTE}$= --0.52$\pm$0.13 for this GC, which is in agreement with the low $\alpha$-element content of this object (\citealt{brown1996}, \citeyear{brown1997}, \citealt{pritzl2005}, \citealt{villanova2013}, \citealt{francois2014}, \citealt{frelijj2021}).

        \item Our S abundance measurement for the star $\#$3416 of Tr\,5 is in agreement within the  errors with the findings of \cite{caffau2014}. We obtained a new and more robust result for S in this OC based on a larger sample of targets. We found an S content for Tr\,5 of about [S/Fe]$_{\rm NTLE}$=0.05$\pm$0.20. 
        
        \item The metallicity and the S content of Tr\,20 are [Fe/H]= 0.06$\pm$0.15 and [S/Fe]$_{\rm NLTE}$= --0.28$\pm$0.21, respectively.

\end{itemize}

\noindent
In summary, our conclusions are that Tr\,5 behaves similarly to the Galactic disk and that Rup\,106 experienced a different nucleosynthesis process from the Galactic halo clusters, supporting the notion that it is an accreted GC that formed outside the MW. Finally, Tr\,20 is an OC belonging and formed in the MW.

\begin{table*}
        \centering
        \caption{Observational dataset.}
        \label{data}
        \begin{tabular}{ccccccccc}      
                \hline\hline
                Star & RA & DEC & G&BP$-$RP&DATE-OBS&S/N&DATE-OBS&S/N\\
                &&&&&R580&at 670 nm&R860$^{(c)}$& at 900 nm\\
                &[h:m:s]&[$^{\circ}$:':'']&[mag]&[mag]&[YYYY-MM-DD]&&[YYYY-MM-DD]&\\
                \hline
                Ruprecht\,106&&&&&&&&\\
                &&&&&&&&\\
                Ru676&12:38:43.83&-51:09:50.9&15.93&1.44&2002 May 07&80&2017 Feb 23&65\\
                Ru801&12:38:36.00&-51:09:37.5&16.24&1.34&2002 May 09&60&2017 Feb 23&80\\
                Ru1614&12:38:36.55&-51:08:25.6&14.10&1.81&2002 May 06&80&2017 Feb 23&130\\
                Ru1863&12:38:34.75&-51:08:01.9&15.97&1.38&2002 May 08&60&2017 Feb 23&70\\
                Ru1951&12:38:35.47&-51:07:52.5&14.97&1.56&2002 May 07&90&2017 Feb 23&80\\
                Ru2004&12:38:40.96&-51:07:46.3&14.27&1.75&2002 May 07&100&2017 Feb 23&130\\
                Ru2032&12:38:48.93&-51:07:43.2&16.11&1.42&2002 May 08&90&2017 Feb 23&60\\
                \hline
                Trumpler\,5&&&&&&&&\\
                &&&&&&&&\\
                1318&6:36:53.4&9:25:34.5&14.59&2.01&2012 Feb 11&40&2017 Oct 03&60\\
                &&&&&&&2017 Nov 13&45\\
                &&&&&&&2017 Dec 29&50\\
                \cline{6-9}
                3416&6:36:40.2&9:29:47.8&14.42&1.89&2012 Feb 11$^{(a)}$&40&2017 Oct 03&40\\
                &&&&&2013 Oct 19$^{(b)}$&103&2017 Nov 13&60\\
                &&&&&&&2017 Dec 29&55\\
                \cline{6-9}
                3678&6:36:23.8&9:30:28.5&14.31&1.88&2012 Feb 11&15&2017 Oct 03&40\\
                &&&&&&&2017 Nov 13&80\\
                &&&&&&&2017 Dec 29&60\\
                \cline{6-9}
                4528&6:36:39.8&9:32:18.6&14.38&1.97&-&-&2017 Oct 03&50\\
                &&&&&&&2017 Nov 13&60\\
                &&&&&&&2017 Dec 29&55\\
                \cline{6-9}
                4791&6:36:33.1&9:33:3.6&14.34&1.92&-&-&2017 Oct 03&35\\
                &&&&&&&2017 Nov 13&65\\
                &&&&&&&2017 Dec 29&40\\
                \cline{6-9}
                4876&6:36:21.7&9:33:23.1&14.09&1.79&-&-&2017 Oct 03&50\\
                &&&&&&&2017 Nov 13&60\\
                &&&&&&&2017 Dec 29&40\\
                \hline
                Trumpler\,20&&&&&&&&\\
                &&&&&&&&\\
                100292&12:39:04.11&-60:34:00.2&12.98&1.74&2013 Mar 25&200&2017 Apr 19&160\\
                340&12:39:15.78&-60:34:40.6&14.25&1.56&2013 Mar 14&90&2017 Apr 19&130\\
                542&12:39:12.01&-60:36:32.1&14.28&1.54&2013 Mar 26&100&2017 Apr 19&80\\
                591&12:40:04.50&-60:36:56.6&13.09&1.85&2013 Feb 20&100&2017 Apr 19&160\\
                \hline
                \hline
                \multicolumn{9}{l}{\small \textbf{Notes.} $^{(a)}$ UVES spectrum. $^{(b)}$ MIKE spectrum. $^{(c)}$ Data collected between the nights: 2017 Feb 23, and 2017 Mar 20 for Rup\,106;}\\
                \multicolumn{9}{l}{\small   2017 Apr 19, and 2017 Apr 24 for Tr\,20.} \\
        \end{tabular}
\end{table*}

\begin{table}   
        \centering
        \caption{Atmospheric parameters and metallicities obtained in this work. Typical errors on the atmospheric parameters are indicated. The errors on the reported [Fe/H] abundances are the standard deviations of the measurements of individual lines.}
        \label{parameters}
        \begin{tabular}{cccccc} 
                \hline\hline
                Star & \teff & \logg & $\xi$ &[Fe/H]\\
                &$\pm50$&$\pm0.1$&$\pm0.1$&\\
                &[K]&[cgs]&[kms$^{-1}$]&&\\
                \hline
                Ruprecht\,106&&&&&\\
                &&&&&\\
                676&4806&1.54&1.91&-1.37 $\pm$ 0.11\\
                801&4881&1.71&1.84&-1.32 $\pm$ 0.13\\
                1614&4314&0.52&2.33&-1.38 $\pm$ 0.12\\
                1863&4820&1.57&1.90&-1.41 $\pm$ 0.12\\
                1951&4554&1.01&2.13&-1.34 $\pm$ 0.11\\
                2004&4353&0.59&2.29&-1.45 $\pm$ 0.11\\
                2032&4838&1.61&1.88&-1.36 $\pm$ 0.13\\
                \hline
                Trumpler\,5&&&&&\\
                &&&&&\\
                1318&4582&2.55&1.04&-0.48 $\pm$ 0.15\\
                3416&4869&2.52&1.27&-0.51 $\pm$ 0.13\\
                3678&4910&2.47&1.33&-0.48 $\pm$ 0.14\\
                4528&4686&2.47&1.16&-0.45 $\pm$ 0.16\\
                4791&4818&2.47&1.26&-0.57 $\pm$ 0.15\\
                4876&5163&2.41&1.56&-0.50 $\pm$ 0.13\\
                \hline
                Trumpler\,20&&&&&\\
                &&&&&\\
                100292&4575&2.16&1.24&0.00 $\pm$ 0.15\\
                340&4997&2.86&1.18&0.04 $\pm$ 0.15\\
                542&5042&2.88&1.19&0.15 $\pm$ 0.14\\
                591&4382&2.18&1.07&0.04 $\pm$ 0.16\\
                \hline
                \hline
        \end{tabular}
\end{table}

\begin{table*}  
        \centering
        \caption{Radial velocities and dates of observation of the stars of our sample.
         The (*) values are those of \cite{monaco2014}.}
        \label{RV}
        \begin{tabular}{ccccc}  
                \hline\hline
                Star & DATE-OBS & RV& DATE-OBS &RV\\
                &R580&R580&R860$^{(c)}$&R860\\
                &&km s$^{-1}$&&km s$^{-1}$\\
                \hline
                Ruprecht\,106&&&&\\
                &&&&\\
                676&2002 May 07&-37.50 $\pm$ 0.22&2017 Feb 23&-37.59 $\pm$ 0.37\\
                801&2002 May 09&-38.87 $\pm$ 0.23&2017 Feb 23&-39.21 $\pm$ 0.42\\
                1614&2002 May 06&-36.28 $\pm$ 0.19&2017 Feb 23&-36.98 $\pm$ 0.22\\
                1863&2002 May 08&-36.99 $\pm$ 0.21&2017 Feb 23&-36.63 $\pm$ 0.41\\
                1951&2002 May 07&-36.43 $\pm$ 0.18&2017 Feb 23&-36.76 $\pm$ 0.19\\
                2004&2002 May 07&-39.01 $\pm$ 0.27&2017 Feb 23&-38.66 $\pm$ 0.24\\
                2032&2002 May 08&-37.47 $\pm$ 0.19&2017 Feb 23&-38.10 $\pm$ 0.36\\
                \hline
                Trumpler\,5&&&&\\
                &&&&\\
                1318&2012 Feb 11&48.10 $\pm$ 0.30$^{(*)}$&2017 Oct 03&48.09 $\pm$ 0.07\\
                &&&2017 Nov 13&48.41 $\pm$ 0.26\\
                &&&2017 Dec 29&48.57 $\pm$ 0.18\\
                \cline{2-5}
                3416&2012 Feb 11$^{(a)}$ &49.80 $\pm$ 0.10$^{(*)}$&2017 Oct 03&49.68 $\pm$ 0.11\\
                &2013 Oct 19$^{(b)}$ &50.50 $\pm$ 0.20$^{(*)}$&2017 Nov 13&50.24 $\pm$ 0.31\\
                &&&2017 Dec 29&50.29 $\pm$ 0.18\\
                \cline{2-5}
                3678&-&-&2017 Oct 03&49.39 $\pm$ 0.11\\
                &&&2017 Nov 13&49.75 $\pm$ 0.32\\
                &&&2017 Dec 29&49.94 $\pm$ 0.18\\
                \cline{2-5}
                4528&-&-&2017 Oct 03&52.41 $\pm$ 0.09\\
                &&&2017 Nov 13&52.77 $\pm$ 0.15\\
                &&&2017 Dec 29&52.99 $\pm$ 0.18\\
                \cline{2-5}
                4791&-&-&2017 Oct 03&49.44 $\pm$ 0.09\\
                &&&2017 Nov 13&50.09 $\pm$ 0.32\\
                &&&2017 Dec 29&49.93 $\pm$ 0.42\\
                \cline{2-5}
                4876&-&-&2017 Oct 03&49.34 $\pm$ 0.09\\
                &&&2017 Nov 13&49.97 $\pm$ 0.41\\
                &&&2017 Dec 29&49.76 $\pm$ 0.18\\
                \hline
                Trumpler\,20&&&&\\
                &&&&\\
                100292&2013 Mar 25&-39.28 $\pm$ 0.14&2017 Apr 19&-39.51 $\pm$ 0.09\\
                340&2013 Mar 14&-40.08 $\pm$ 0.15&2017 Apr 19&-39.80 $\pm$ 0.14\\
                542&2013 Mar 26&-39.38 $\pm$ 0.14&2017 Apr 19&-40.61 $\pm$ 0.15\\
                591&2013 Feb 20&-40.20 $\pm$ 0.35&2017 Apr 19&-40.95 $\pm$ 0.19\\
                \hline
                \hline
                \multicolumn{5}{l}{\small \textbf{Notes.} $^{(a)}$ UVES spectrum. $^{(b)}$ MIKE spectrum. $^{(c)}$ Data collected between the nights:}\\
                \multicolumn{5}{l}{\small   2017 Feb 23, and 2017 Mar 20 for Rup\,106; 2017 Apr 19, and 2017 Apr 24 for Tr\,20.} \\
        \end{tabular}
\end{table*}

\begin{table}
        \centering
        \caption{Atomic parameters of the sulfur lines.}
        \label{par_ato}
        \small
        \centering
        \begin{tabular}{ccccc}  
                \hline\hline
                Wavelength & Mult. & Transition & log $gf$ & $\chi_{\rm    low}$\\
                $[$\AA$]$& & & & $[$eV$]$\\
                \hline
                6757.153 & 8 &$^5$P$_3-^5$D$_4^{\rm o}$ &-0.35&9.704\\          
                8694.709 & 6 &$^5$P$_3-^5$D$_4^{\rm o}$ &0.05&9.295\\
                9212.863 & 1 &$^5$S$_2^{\rm o}-^5$P$_3$ &0.40 &6.525\\
                9228.093 & 1 &$^5$S$_2^{\rm o}-^5$P$_2$ &0.25&6.525\\
                9237.538 & 1 &$^5$S$_2^{\rm o}-^5$P$_1$ &0.03&6.525\\
                \hline
                \hline
        \end{tabular}
\end{table}

\begin{table*}  
        \centering
        \caption{LTE and NLTE sulfur abundances obtained line by line for the stars analyzed in this work. The A(S) obtained from line 6757 $\AA$ are included in the mean LTE and NLTE <A(S)> values.}
        \label{lines}
        \begin{tabular}{cccccccccccc}   
                \hline\hline
                Star & 6757 & 8694& 9212 & 9228 & 9237 & <A(S)>& 8694 & 9212 & 9228 & 9237 &<A(S)> \\
                &&LTE & LTE & LTE & LTE & LTE& NLTE & NLTE & NLTE & NLTE&NLTE\\
                \hline
                Ruprecht\,106&&&&&&&&&&&\\
                &&&&&&&&&&&\\
                676&-&-&5.38&5.37&-&5.38 $\pm$ 0.01&-&5.17&5.23&-&5.20 $\pm$ 0.04\\
                801&-&-&5.63&5.60&-&5.62 $\pm$ 0.02&-&5.36&5.39&-&5.38 $\pm$ 0.03\\
                1614&-&-&5.36&5.17&-&5.27 $\pm$ 0.13&-&5.31&5.24&-&5.28 $\pm$ 0.05\\
                1863&-&-&5.51&5.52&-&5.52 $\pm$ 0.01&-&5.26&5.32&-&5.29 $\pm$ 0.04\\
                1951&-&-&5.40&5.41&-&5.41 $\pm$ 0.01&-&5.23&5.31&-&5.27 $\pm$ 0.06\\
                2004&-&-&5.29&5.10&-&5.19 $\pm$ 0.13&-&5.22&5.14&-&5.18 $\pm$ 0.05\\
                2032&-&-&5.40&5.59&-&5.49 $\pm$ 0.14&-&5.18&5.38&-&5.28 $\pm$ 0.14\\
                \hline
                Trumpler\,5&&&&&&&&&&&\\
                &&&&&&&&&&&\\
                1318&7.06&6.87&7.05&-&-&6.99 $\pm$ 0.11&6.85&6.80&-&-&6.90 $\pm$ 0.14\\
                3416&7.05&6.85&7.04&-&-&6.98 $\pm$ 0.11&6.81&6.67&-&-&6.84 $\pm$ 0.19\\
                3678&6.88&6.85&6.85&-&-&6.86 $\pm$ 0.02&6.81&6.47&-&-&6.72 $\pm$ 0.22\\
                4528&-&6.71&6.88&-&-&6.79 $\pm$ 0.12&6.68&6.59&-&-&6.64 $\pm$ 0.06\\
                4791&-&6.98&7.17&-&-&7.08 $\pm$ 0.13&6.94&6.81&-&-&6.88 $\pm$ 0.09\\
                4876&-&6.44&6.63&-&-&6.54 $\pm$ 0.13&6.39&6.16&-&-&6.28 $\pm$ 0.16\\
                \hline
                Trumpler\,20&&&&&&&&&&&\\
                &&&&&&&&&&&\\
                100292&7.15&-&7.13&-&7.14&7.14 $\pm$ 0.01&-&6.80&-&6.88&6.94 $\pm$ 0.18\\
                340&6.98&6.99&6.98&-&6.98&6.98 $\pm$ 0.01&6.95&6.68&-&6.75&6.84 $\pm$ 0.15\\
                542&7.10&7.10&7.08&-&7.09&7.09 $\pm$ 0.01&7.05&6.77&-&6.85&6.94 $\pm$ 0.16\\
                591&7.17&-&7.17&-&7.18&7.17 $\pm$ 0.01&-&6.91&-&6.99&7.02 $\pm$ 0.13\\
                \hline
                \hline
        \end{tabular}
\end{table*}

\newpage

\newpage

\begin{acknowledgements}
Support for the author FL is provided by CONICYT-PFCHA/Doctorado Nacional año 2020-folio 21200677.
EC and PB acknowledge support from the Agence Nationale de la Recherche (ANR project
ANR-18-CE31-0017). S.V. gratefully acknowledges the support provided by Fondecyt reg. 1220264, the ANID BASAL projects 
 ACE210002 and  FB210003.
This work has made use of data from the European Space Agency (ESA) mission
{\it Gaia} (\url{https://www.cosmos.esa.int/gaia}), processed by the {\it Gaia}
Data Processing and Analysis Consortium (DPAC,
\url{https://www.cosmos.esa.int/web/gaia/dpac/consortium}). Funding for the DPAC
has been provided by national institutions, in particular the institutions
participating in the {\it Gaia} Multilateral Agreement.
The Gaia-ESO Survey is based on observations made with ESO Telescopes at the La Silla or Paranal Observatories under programme ID(s) 072.D-0019(B), 072.D-0309(A), 072.D-0337(A), 072.D-0406(A), 072.D-0507(A), 072.D-0742(A), 072.D-0777(A), 073.C-0251(B), 073.C-0251(C), 073.C-0251(D), 073.C-0251(E), 073.C-0251(F), 073.D-0100(A), 073.D-0211(A), 073.D-0550(A), 073.D-0695(A), 073.D-0760(A), 074.D-0571(A), 075.C-0245(A), 075.C-0245(C), 075.C-0245(D), 075.C-0245(E), 075.C-0245(F), 075.C-0256(A), 075.D-0492(A), 076.B-0263(A), 076.D-0220(A), 077.C-0655(A), 077.D-0246(A), 077.D-0484(A), 078.D-0825(A), 078.D-0825(B), 078.D-0825(C), 079.B-0721(A), 079.D-0178(A), 079.D-0645(A), 079.D-0674(A), 079.D-0674(B), 079.D-0674(C), 079.D-0825(B), 079.D-0825(C), 079.D-0825(D), 080.B-0489(A), 080.B-0784(A), 080.C-0718(A), 081.D-0253(A), 081.D-0287(A), 082.D-0726(A), 083.B-0083(A), 083.D-0208(A), 083.D-0671(A), 083.D-0682(A), 083.D-0798(B), 084.D-0470(A), 084.D-0693(A), 084.D-0933(A), 085.D-0205(A), 086.D-0141(A), 087.D-0203(B), 087.D-0230(A), 087.D-0276(A), 088.B-0403(A), 088.B-0492(A), 088.C-0239(A), 088.D-0026(A), 088.D-0026(B), 088.D-0026(C), 088.D-0026(D), 088.D-0045(A), 089.D-0038(A), 089.D-0298(A), 089.D-0579(A), 090.D-0487(A), 091.D-0427(A), 092.D-0171(C), 092.D-0477(A), 093.D-0286(A), 093.D-0818(A), 094.D-0363(A), 094.D-0455(A), 171.D-0237(A), 187.B-0909(A), 188.B-3002(A), 188.B-3002(B), 188.B-3002(C), 188.B-3002(D).
\end{acknowledgements}
\bibliographystyle{aa.bst}
\bibliography{biblio}
\newpage
\appendix
\section{ATLAS 9 versus ATLAS 12 models}
\label{app_9_12}

\begin{figure}[h]
\centering
\resizebox{7.1cm}{!}{\includegraphics{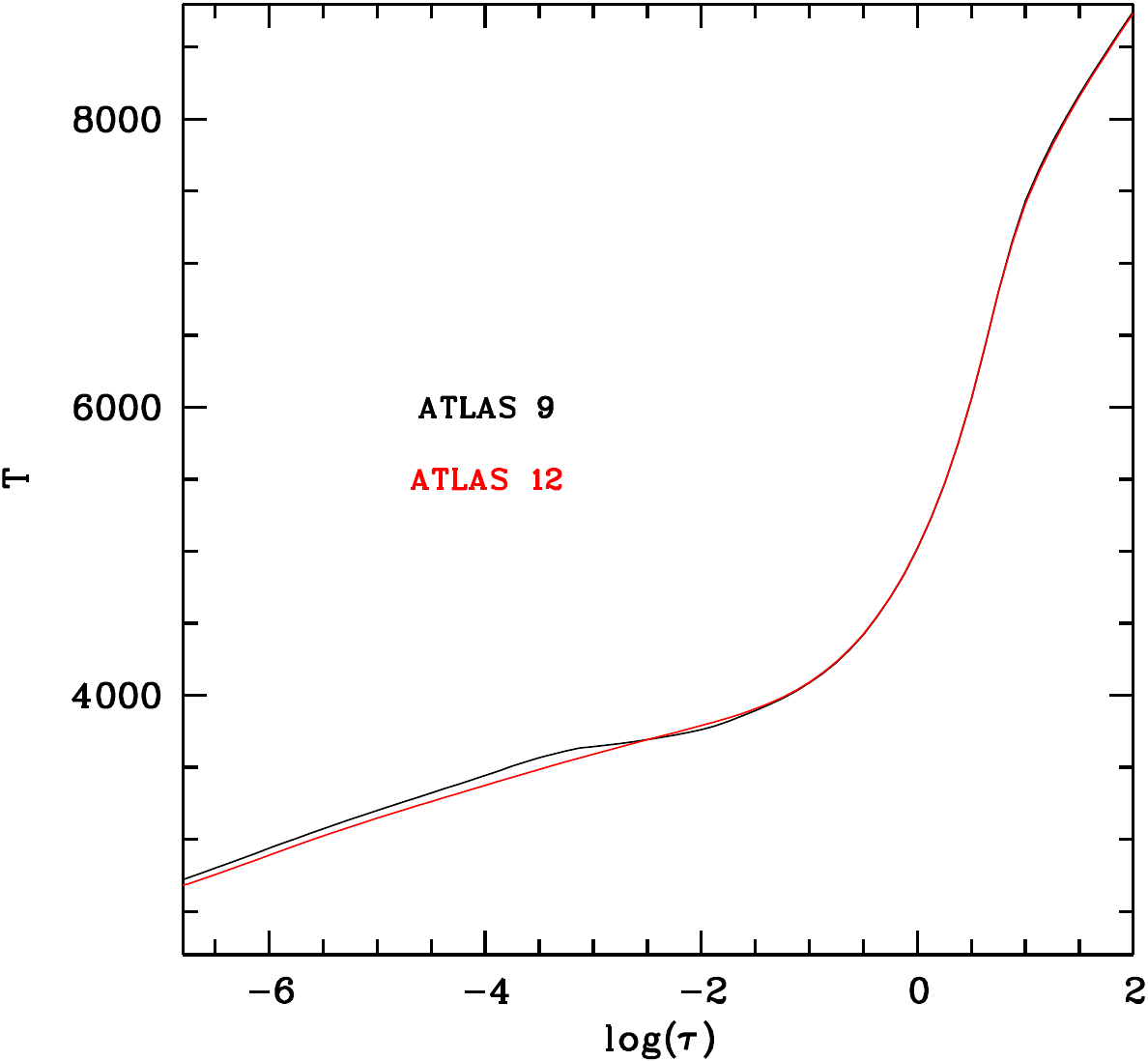}}
\caption{Comparison of the temperature structure of
an ATLAS 9 (black) and an ATLAS 12 model with \teff = 4582,
log g = 2.55, metallicity -0.5, and microturbulence 2\,\kms.\label{plmod}}
\end{figure}

The comparison of these two classes of models
has been made several times; we here refer the reader
to section 9 of \cite{Castelli05}. In that paper, the algorithm of opacity 
sampling is described and details on how to run
ATLAS 12 are provided. 
It has to be understood that the two versions of 
ATLAS codes are not identical, meaning that some 
small differences between the models are to be expected.
In Fig.\,\ref{plmod} we compare the temperature
structures of two models appropriate for the study
of star \#\,1318 in  \relax Trumpler\,5.
Up to log($\tau$) = --2.8 the two models
are identical, for the outer layers the ATLAS 12 model
is about 70\,K cooler. For all lines that form deeper
than log($\tau$) = --2.8, that is the case of the sulfur
lines, the two models will provide identical results. 
A comparison of the accuracy or strength of the two classes of models has not been carried out. 
Computationally, the ATLAS 12 model takes over 30 times longer
than the ATLAS 9 model.

\end{document}